\documentclass{optica-article}

\journal{opticajournal} 

\articletype{Research Article}

\usepackage{lineno}
\usepackage{orcidlink}

\usepackage{caption}
\usepackage{subcaption}
\usepackage{hyperref}

\newcommand{\p}[1]{\ensuremath{\left( #1 \right)}}

\usepackage[normalem]{ulem}
\usepackage{cancel}

\begin{document}

\title{Influence of optical aberrations on the accuracy of an atomic gravimeter}

\author{\orcidlink{0009-0008-4264-7511} Louis Pagot, \orcidlink{0000-0002-4746-2400} S\'ebastien Merlet, and \orcidlink{0000-0003-0659-5028} Franck Pereira Dos Santos\authormark{*}}

\address{LNE-SYRTE, Observatoire de Paris, Université PSL, CNRS, Sorbonne Université, 61 Avenue de l'Observatoire, 75014 Paris, France}

\email{\authormark{*}franck.pereira@obspm.fr} 


\begin{abstract*} 
We present numerical simulations of the impact of laser beam wavefront aberrations in cold atom interferometers. We demonstrate that to reach accuracy at the mrad level, it is insufficient to base simulations on a description of retroreflective optics using only low-order Zernike polynomials, as the results will then depend on the decomposition order and the decomposition technique chosen. Moreover, simulations with high-order Zernike polynomials or equivalently high spatial frequency components require the propagation of aberrations to be taken into account, rather than adding them to the ideally propagated beam. Finally, we examine the impact of the parameters of the atomic source and show that the use of delta-kicked atomic cloud would efficiently mitigate the impact of this systematic effect.

\end{abstract*}

\section{Introduction}
 Since the first realizations of a gyroscope \cite{riehleOpticalRamseySpectroscopy1991} and a gravimeter \cite{kasevichAtomicInterferometry1991} with atom interferometers, significant research efforts have been made to tailor these sensors for a variety of applications \cite{bongsTakingAIQuantumSensors2019}, and to improve both their sensitivity and accuracy \cite{geigerHighAccuracyInertialMeasurements2020}. Atomic interferometers allow to measure inertial quantities such as accelerations, like gravity, gravity gradients, and rotations, in laboratory setups \cite{riehleOpticalRamseySpectroscopy1991,kasevichAtomicInterferometry1991,louchet-chauvetInfluenceTransverseMotion2011,GautierAccurateMeasurementSagnac2022,janvierCompactDifferentialGravimeter2022}, on dynamic platforms \cite{geigerAirborneMatterWave2011,bidelMarineGravimetry2018} or on field \cite{antoni-micollierDetectingVolcano2022,gunterMobileFieldMeasurements2024}. They allow for performing ultrasensitive tests of fundamental physics \cite{bouchendiraNewDeterminationFineConstant2011, ballandQuectonewtonLocalForceSensor2023} on the ground, and first cold atom experiments in space \cite{BeckerSpaceBornBEC2018,AvelineObservingBECOrbit2020,liRealizationCAGyro2024} are paving the way for ambitious missions operating these sensors onboard satellites \cite{levequeCarioqaDefinitionQuantumPathfinder2022,abendRecentDevelopmentsQT2023,struckmannPlatformEnvironmentRequirements2024}.
 
 Matter-wave interferometers consist in a series of coherent light pulses, separated by free evolution time, which split, deflect and finally recombine the atomic wavepackets. At each light-matter interaction, the phase of the light field is imprinted on the wavepacket. As times and laser frequencies are well defined, so does the scale factor of the interferometric phase. In addition, many of the systematic effects can be theoretically modeled and then suppressed thanks to sequences of measurements \cite{louchet-chauvetInfluenceTransverseMotion2011}. Among the remaining systematics, one of the most difficult to precisely evaluate is the one related to the wavefront aberrations of the lasers \cite{geigerHighAccuracyInertialMeasurements2020}, which perturb the laser phases imprinted at the time of the pulses, depending on the position of the wavepacket in the light field, \cite{louchet-chauvetInfluenceTransverseMotion2011} and modify the trajectories of the atoms \cite{badeObservationExtraPhotonRecoil2018,cervantesEffectAperture2024}. The phase shift resulting from wavefront aberrations has been studied through different approaches in existing apparatus. For instance, the size of the detection \cite{schkolnikEffectWavefrontAberrations2015} or the size of the Raman beams \cite{badeObservationExtraPhotonRecoil2018,cervantesEffectAperture2024,zhouObservingEffectWavefrontAberrations2016} have been modulated, as well as the temperature of the atomic cloud \cite{karcherImprovingAccuracyAtom2018}. Moreover, it has been demonstrated that the bias caused by the introduction of a characterized optics could be either simulated and retrieved from the measured value \cite{schkolnikEffectWavefrontAberrations2015} or corrected thanks to a deformable mirror \cite{trimecheActiveControlLaser2017}. This systematic effect will also be the subject of future studies for specifications in the context of state-of-the-art atom interferometers aboard space missions \cite{levequeCarioqaDefinitionQuantumPathfinder2022,abendRecentDevelopmentsQT2023}.
 
 In this paper, we demonstrate that to accurately calculate the phase bias to better than the mrad level, we cannot restrict the description of retro-reflecting optics to low order Zernike polynomials \cite{Zernike_low_order}. In addition, we show that considering aberrations described by high Zernike orders, or equivalently with high spatial frequencies, imposes to take their propagation into account. We apply these results to the concrete case of a gravimeter experiment \cite{louchet-chauvetInfluenceTransverseMotion2011} with characterized, high-quality retro-reflecting mirrors.

\section{Description of the gravimeter}

The atom interferometer considered here is a gravimeter. It is a two-wave interferometer based on a $\frac{\pi}{2} - \pi - \frac{\pi}{2}$ sequence of stimulated Raman transitions separated by a free evolution time $T = 80$~ms. Thanks to a sequence of four measurements, most systematic shifts are eliminated \cite{louchet-chauvetInfluenceTransverseMotion2011}, the remaining effects being the Coriolis phase shift, which can be evaluated by rotating the gravimeter of $180^{\circ}$ or suppressed by compensating the Earth rotation \cite{Lan2012Influence}, and the one resulting from wavefront aberrations. For free-falling atoms in the gravity field $g$ and Raman counter-propagating laser fields aligned along the vertical axis with a linear sweep $\alpha$ of their angular frequency difference, the interferometer phase $\Delta \Phi$ is given by 
\begin{equation}\label{eq0:InterfPhase}
    \Delta \Phi = \phi_1 - 2\phi_2 + \phi_3 = \p{k_{\text{eff}}g - \alpha} T^2 + \Delta \varphi.
\end{equation}
$k_{\text{eff}} \approx \frac{4\pi}{\lambda}$ is the effective wavevector of the Raman transition, $\phi_i$ is the phase difference of the laser beams at the position of the \textit{i}$^\text{th}$ Raman pulse, and $\Delta \varphi$ is the interferometer phase shift caused by wavefront aberrations. Since the sensitivity factor is $k_{\text{eff}}T^2 = 10^{5}$~rad~m$^{-1}$~s$^2$, a phase shit of $1$~mrad corresponds to a bias of $10^{-8}$~m~s$^{-2} = 1$~$\mu$Gal in the measured gravity value. To assess the contribution of this effect and eventually subtract it from the experimental measurements, the numerical simulation presented thereafter is used.

The simulation consists of a Riemann integral over the 5D phase space of the atomic cloud, with the initial distribution in longitudinal position disregarded. Assuming initial normal distributions in each dimension, we define the finite radius of the hyper-sphere on which to integrate by fixing the weight of the distribution that will be neglected. 
Choosing a weight of $0.25$~\% leads to a bias lower than $5 \times 10^{-2}$~$\mu$Gal in the simulations. 
The field amplitude corresponds to a Gaussian distribution of $12$~mm waist, the radius at $1/e^2$ intensity. It is defined at the collimator output and propagated to the positions of the four laser pulses used in the experiment, one for the velocity selection and three for the Raman pulses, by solving the Helmholtz equation with Fourier transforms. These positions correspond to the classical path of the free-falling atoms despite the separation of the different momentum components of the wavepackets. This approximation is valid as long as the phase of the laser field evolves linearly along the separation distance, which is the case for low enough transverse spatial frequencies of the aberrations as shown below. The propagation length with the retro-reflection is roughly $1$~m, thus the field is represented on a square grid with $84$~mm sides and $2048$ points along each transverse dimension to satisfy the sampling condition \cite{kozackiNumericalErrorsDiffraction2008}. The transition probability is calculated for multiple sweep frequencies $\alpha$, and eventually a fit with the function $P = P_m - \frac{C}{2} \cos{\p{\Delta \Phi}}$ allows to extract the interferometer phase bias \eqref{eq0:InterfPhase} together with the contrast $C$ and the mean value $P_m$ of the transition probability. With the typical parameters chosen to run the simulations, the uncertainty obtained on the phase from the fit is typically $6 \times 10^{-3}$~$\mu$Gal. Introducing deviations from the free falling trajectories due to extra-photon recoil \cite{badeObservationExtraPhotonRecoil2018} modify the results by quantities that are typically orders of magnitude smaller than the results without taking it into account, which is consistent with the typical size of the cloud, and the spatial frequencies and amplitudes of aberrations considered \cite{cervantesEffectAperture2024}. Hence, for simplicity and comparison with analytical models the results shown below do not take extra-photon recoil into account.

\section{Mirror with an arbitrary surface}\label{MirrorArbitrary}

To our knowledge, most studies of the effect of laser wavefront distortions caused by retro-reflecting optics in atomic interferometers have used low-order Zernike polynomials to describe the wavefront aberrations \cite{schkolnikEffectWavefrontAberrations2015, karcherImprovingAccuracyAtom2018, trimecheActiveControlLaser2017}, as these polynomials are generally used for this purpose in optics \cite{bornwolfPrinciplesOptics2019, niuZernikePolynomialsApplications2022} and in particular for the analysis of Shack-Hartmann sensor and optical interferometers measurements. Though the contribution of high-order Zernike polynomials is not necessarily zero, even for large atomic clouds \cite{louchet-chauvetInfluenceTransverseMotion2011}. Moreover, propagation is usually not taken into account: aberrations are simply added to the wavefront of the propagated beam for the ideal case of a perfectly flat mirror \cite{zhouObservingEffectWavefrontAberrations2016,karcherImprovingAccuracyAtom2018}.

\begin{figure}[ht!]
     \centering
     \begin{subfigure}[b]{0.45\textwidth}
         \centering
         \includegraphics[height=5.1cm]{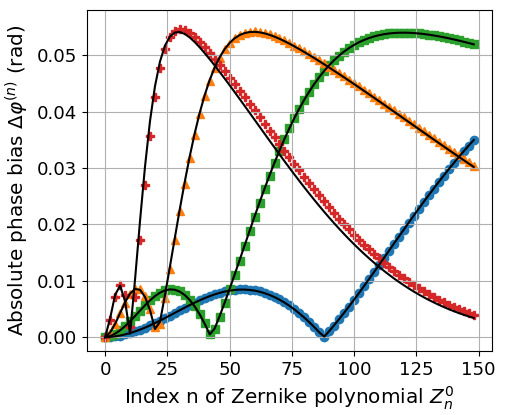}
         \caption{Addition of aberrations}
         \label{Fig1a:DipolarTrap_add}
     \end{subfigure}
     \begin{subfigure}[b]{0.41\textwidth}
         \centering
         \includegraphics[height=5.1cm]{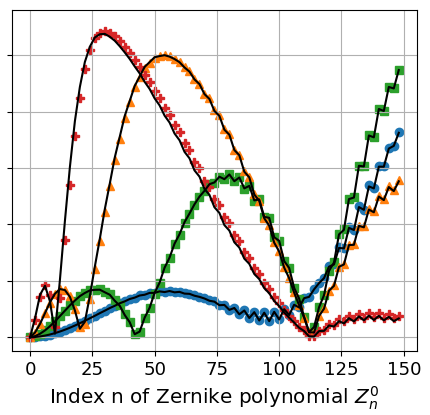}
         \caption{Propagation of aberrations}
         \label{Fig1b:DipolarTrap_prop}
     \end{subfigure}
     \begin{subfigure}[b]{0.04\textwidth}
         \centering
         \raisebox{2.5cm}{\includegraphics[width=1.8cm]{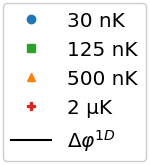}}
     \end{subfigure}
     \caption{Absolute phase bias caused by a mirror surface represented by a Zernike polynomial $Z_n^0$ on a $2R = 27$~mm diameter with an amplitude $\gamma = \frac{\lambda}{200}$, for an interferometer duration $2T=160$~ms, cloud size $\sigma_{xy} = 10$~$\mu$m and temperature $\Theta \in \left\{30, 125, 500, 2000 \right\}$~nK. Markers corresponds to simulations where the aberrations have been: (a) added to the beam, (b) propagated with the beam. The continuous black lines $\Delta \varphi^{1\text{D}}$ correspond to the 1D integral: (a) for addition \eqref{eq1:WFAddition}, (b) for propagation \eqref{eq1:WFPropagation} of the aberrations. Errorbars are smaller than the markers.}
     \label{Fig1:DipolarSimulation}
\end{figure}

To examine the effect of each Zernike polynomial, we start by considering mirrors with an arbitrary surface defined by Zernike polynomials with rotational invariance $Z_n^0$ with $n$ even, as in the case of an atomic cloud with the same symmetry property and coaxially centered, the contribution of the others Zernike polynomials, with at least one axis of anti-symmetry, is null. When the Zernike polynomial defined on a disk of radius $R$ is added to the beam phase, the amplitude outside this disk is set to zero. In Figure~\ref{Fig1:DipolarSimulation}, the phase shift of the gravimeter is simulated with an atom source evaporatively cooled in a dipole trap, resulting in an atomic cloud of size of order of $\sigma_{xy} = 10$~$\mu$m and the possibility of adjusting the temperature \cite{louchet-chauvetInfluenceTransverseMotion2011, karcherImprovingAccuracyAtom2018}. Irrespective of the index $n$ of the Zernike polynomial, the amplitude of the aberration is fixed at $\frac{\lambda}{200}$, although the decomposition of mirrors surface lead to a scaling law in $n^{-1}$ for the amplitude of the coefficients, as shown in Figure~\ref{Fig2c:ScallingRotCoeffs} . To consider only the impact of the mirror aberrations, the contribution due to the Gaussian field propagation, obtained with $Z_0^0$, is subtracted. Finally, as the Zernike polynomials $Z_n^0$ have an overall factor $\p{-1}^\frac{n}{2}$, the absolute value of the phase shift is considered for simplicity.

Simulations are carried out for a mirror with radius $R = 13.5$~mm, which corresponds to the largest radius used to analyse the surface of one of the mirrors shown in Figure~\ref{Fig2:SurfaceMirror}, and for different atomic cloud temperatures $\Theta$ from $30$~nK to $2$~$\mu$K. Firstly, mirror surface aberrations are added to the propagated field in the ideal case \cite{zhouObservingEffectWavefrontAberrations2016, karcherImprovingAccuracyAtom2018}, which corresponds to the markers in Figure~\ref{Fig1a:DipolarTrap_add}. These results show a maxima for high-order Zernike polynomials and are similar up to a temperature-dependent scaling factor on the Zernike index $n$, detailed hereafter. Secondly, the mirror aberrations are added to the field at the mirror position and then propagated, corresponding to the markers in Figure~\ref{Fig1b:DipolarTrap_prop}. These second simulations exhibit significant differences at large Zernike orders when the ratio $\frac{\p{n+1}^2 \Delta z_{1,\text{mir}}}{k_{\text{eff}}R^2}$ starts to be of the order of $1$, with $\Delta z_{1,\text{mir}}$ the distance between the mirror and the position of the first Raman pulse, as explained below. Considering that the atomic cloud is initially distributed according to a centered normal distribution of standard deviation $\sigma_{xy}$ for its transverse position and according to a centered normal distribution of standard deviation $\sigma_v = \sqrt{\frac{k_B \Theta}{m_{\text{Rb}}}}$ for its transverse velocity, the atomic distribution along the radial coordinate $\rho$ after a time $t$ is given by
\begin{equation}\label{eq1:Distribution}
    w\p{\rho, t}d\rho = \frac{\rho}{\sigma_\rho^2\p{t}} e^{-\frac{\rho^2}{2\sigma_\rho^2\p{t}}}d\rho \text{ with } \sigma_\rho^2\p{t} = \sigma_{xy}^2 + \sigma_v^2 t^2.
\end{equation}
For a mirror of radius $R$ with a surface described by the Zernike polynomial $Z_n^0$ of amplitude $\gamma$, in the case of added aberrations to an ideally propagated plane wave, the interferometric phase bias is
\begin{equation}\label{eq1:WFAddition}
    \Delta \varphi^{\p{n}}_{\text{add}} =  \int \frac{4 \pi \gamma}{\lambda} Z_n^0\left(\frac{\rho}{R} \right) \left[w\p{\rho, t_0} - 2w\p{\rho, t_0 + T} + w\p{\rho, t_0 + 2T}\right] d\rho.
\end{equation}
$t_0$ is the time between the beginning of the free fall and the first Raman pulse. This model based on a 1D integral corresponds to the continuous black lines in Figure~\ref{Fig1a:DipolarTrap_add} and can be solved analytically by replacing the Zernike polynomial by its Bessel function approximation \eqref{eqA:BesselApprox}
\begin{equation}\label{eq1:ApproxAddition}
    \Delta \varphi^{\p{n}}_{\text{add}} \approx  \frac{4 \pi \gamma_n}{\lambda} e^{-\frac{\sigma_\rho^2\p{t_0}}{2\rho_n^2}} \left[1 - 2 e^{-\frac{\sigma_{v}^2 \p{T^2 + 2t_0 T}}{2\rho_n^2}} + e^{-\frac{\sigma_{v}^2 \p{4T^2 + 4t_0 T}}{2\rho_n^2}} \right].
\end{equation}
$\gamma_n = \p{-1}^{\frac{n}{2}} \gamma$, $\rho_n = \frac{R}{n+1}$ is the characteristic length of the oscillation of the Zernike polynomial close to the center of the mirror. Neglecting the initial size $\sigma_{xy}$ with respect to the expansion term $\sigma_{v} t_0$ in equation \eqref{eq1:ApproxAddition} results in an expression that depends on the product $\p{n+1} \sigma_v$, which explains the scaling feature in the index $n$ with respect to temperature shown in Figure~\ref{Fig1a:DipolarTrap_add}. In addition, averaging the aberrations over the cloud suppresses spatial frequencies greater than the inverse of the atomic cloud size at the first Raman pulse $\sigma_{\rho}^{-1}\p{t_0} $ and is more significant than the $n^{-1}$ typical decrease in amplitude coefficients shown in Figure~\ref{Fig2c:ScallingRotCoeffs}. When decreasing the temperature the bracketed term tends to zero, such that the unbiased $g$ value is obtained \cite{louchet-chauvetInfluenceTransverseMotion2011}. The general expression for the phase of the interferometer \eqref{eq0:InterfPhase} depends on the phase of $U\p{\rho, z_i}$ the product of the complex amplitude of the counter-propagating laser fields at the $i^{\text{th}}$ Raman pulse.
\begin{equation}\label{eq1:PhaseGeneral}
    \begin{split}
        \Delta \Phi = \int [&\text{arg}\p{U\p{\rho, z_1}} w\p{\rho, t_0} - 2\text{arg}\p{U\p{\rho, z_2}}w\p{\rho, t_0 + T} \\ & + \text{arg}\p{U\p{\rho, z_3}} w\p{\rho, t_0 + 2T}] d\rho
    \end{split}
\end{equation}
The phase of $U\p{\rho, z_i}$ is at first order $k_\text{eff} z_i$, where, for simplicity, the frequency sweep is not taken into account. The main contribution to the phase of the interferometer is then $k_\text{eff} g T^2$, which must be subtracted to compare the bias caused by the propagation of the aberrations with that obtained by adding them \eqref{eq1:WFAddition}.
\begin{equation}\label{eq1:WFPropagation}
    \Delta \varphi^{\p{n}}_{\text{prop}} = \Delta \Phi^{\p{n}}_{\text{prop}} - k_\text{eff} g T^2
\end{equation}
$\Delta \Phi^{\p{n}}_{\text{prop}}$ is the interferometer phase obtained for a mirror surface described by a Zernike polynomial $Z_n^0$. This model together with the approximation up to second order of the propagation of the aberrations presented in the \textbf{\nameref{Appendix}}, corresponds to the black lines in Figure~\ref{Fig1b:DipolarTrap_prop}. Using the first order correction \eqref{eqA:FieldApprox} of the propagation of the field with a Zernike wavefront aberration $Z_n^0$ and the same approximations as the ones leading to equation \eqref{eq1:ApproxAddition} results in
\begin{equation}\label{eq1:ApproxPropagation}
    \begin{split}
        \Delta \varphi^{\p{n}}_{\text{prop}} \approx \frac{4 \pi \gamma_n}{\lambda} e^{-\frac{\sigma_\rho^2\p{t_0}}{2\rho_n^2}} &\left[\cos{\p{\frac{\Delta z_{1,\text{mir}}}{k_\text{eff} \rho_n^2}}} - 2 \cos{\p{\frac{\Delta z_{2,\text{mir}}}{k_\text{eff} \rho_n^2}}} e^{-\frac{\sigma_{v}^2 \p{T^2 + 2t_0 T}}{2\rho_n^2}}  \right. \\
        &\left. \ \ + \cos{\p{\frac{\Delta z_{3,\text{mir}}}{k_\text{eff} \rho_n^2}}} e^{-\frac{\sigma_{v}^2 \p{4T^2 + 4t_0 T}}{2\rho_n^2}}\right].
    \end{split}
\end{equation}
$\Delta z_{i,\text{mir}}$ is the distance between the position of the \textit{i}$^\text{th}$ Raman pulse and the mirror. This expression shows a characteristic variation length $l_n = k_\text{eff} \rho_n^2 = \frac{k_\text{eff} R^2}{\p{n+1}^2}$. Consequently, propagation is negligible when $n \ll \sqrt{\frac{2kR^2}{\Delta z_{1,\text{mir}}}}$, which corresponds in the gravimeter configuration to $n \ll 80$. Furthermore, as long as $n \ll \sqrt{\frac{2kR^2}{\Delta z}} \approx 1000$ corresponding to aberrations of typical variation length of $\approx 10$~$\mu$m, where $\Delta z \approx 3$~mm is the maximal longitudinal separation of the atomic wavepackets, the longitudinal position distribution of the cloud and the separation of the components of the wavepackets can be neglected. Finally, when decreasing the temperature, the expression in brackets in equation \eqref{eq1:ApproxPropagation} does not generally tends to zero, unlike the case where aberrations are added \eqref{eq1:ApproxAddition}. To reduce the bias on $g$, it is necessary either to obtain a better mirror with smaller aberrations of amplitude $\gamma$, or to use an atomic cloud with a larger initial size $\sigma_{xy}$ which could be done through delta-kick collimation \cite{kovachyMatterWaveLensing2015}. Note that the analytical model developed here, generalized to all Zernike polynomials and to non-centered atomic clouds \cite{GeneralizationPhaseBias}, could help define specifications on the amplitudes $\gamma$ for retro-reflective optics in particular cold-atom interferometer experiments.

\section{Specific mirror surface}
\begin{figure}[ht!]
     \centering
     \begin{subfigure}[b]{0.48\textwidth}
         \centering
         \includegraphics[width=\textwidth]{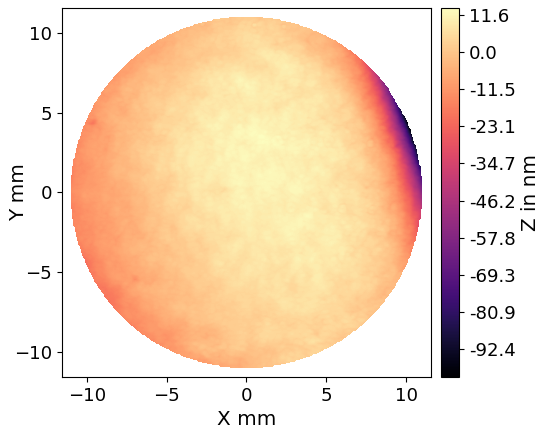}
         \caption{Mirror surface $1$}
         \label{Fig2a:SurfaceMirror1}
     \end{subfigure}
     \begin{subfigure}[b]{0.48\textwidth}
         \centering
         \includegraphics[width=\textwidth]{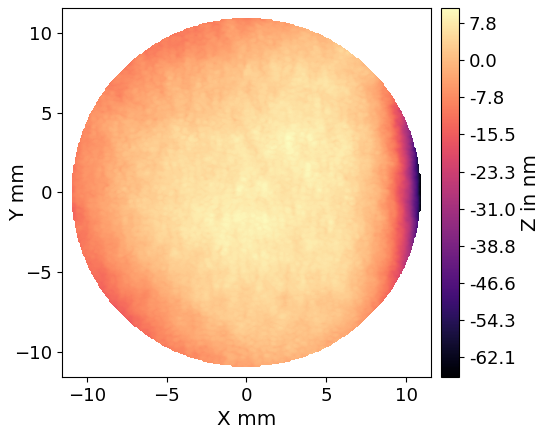}
         \caption{Mirror surface $2$}
         \label{Fig2b:SurfaceMirror2}
     \end{subfigure}
     \hspace{0.4cm}
     \begin{subfigure}[b]{0.60\textwidth}
         \centering
         \includegraphics[width=\textwidth]{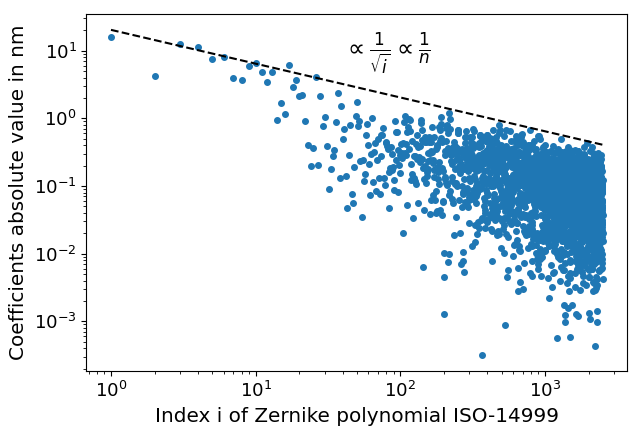}
         \caption{Absolute values of Zernike coefficients of mirror surface~$1$.}
         \label{Fig2c:ScallingRotCoeffs}
     \end{subfigure}
     \caption{Surface of available mirrors, which could replace the current retro-reflecting mirror in the cold atom gravimeter, truncated to $2R = 22$~mm diameter \cite{MirrorSurface_1_2}. The manufacturing specifications on the central $5$~mm radius disk: (a) mirror surface $1$, peak-to-valley amplitude of $9.5$~nm and RMS value of $ 1.5$~nm, (b) mirror surface $2$ and peak-to-valley amplitude of $9.2$~nm, RMS value of $ 1.1$~nm. (c) The coefficients of the Zernike polynomials are calculated with a scalar product decomposition of the raw data of mirror surface~$1$ up to $Z_{98}^0$ with ISO-14999 index $2499$ \cite{ISO_14999}. The envelope of Zernike coefficients for mirror surface~$2$ follows the same scaling law $i^{-\frac{1}{2}}$ ($\propto n^{-1}$ for $m=0$), but for reasons of visibility is not shown.}
     \label{Fig2:SurfaceMirror}
\end{figure}

\begin{figure}[ht!]
     \centering
     \begin{subfigure}[b]{0.48\textwidth}
         \centering
         \includegraphics[width=\textwidth]{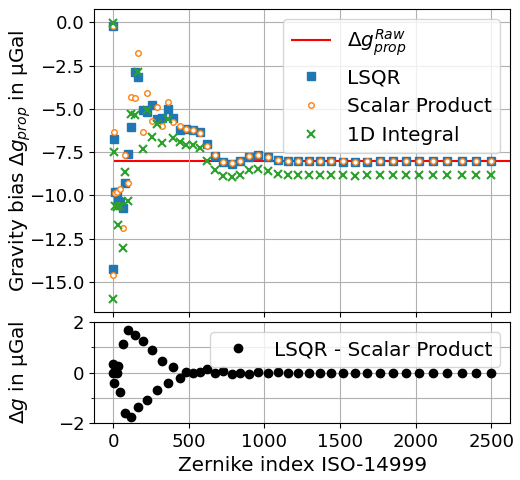}
         \caption{Mirror aberrations propagated with the field}
         \label{Fig3a:DecompositionSN1_11mm}
     \end{subfigure}
     \begin{subfigure}[b]{0.48\textwidth}
         \centering
         \includegraphics[width=\textwidth]{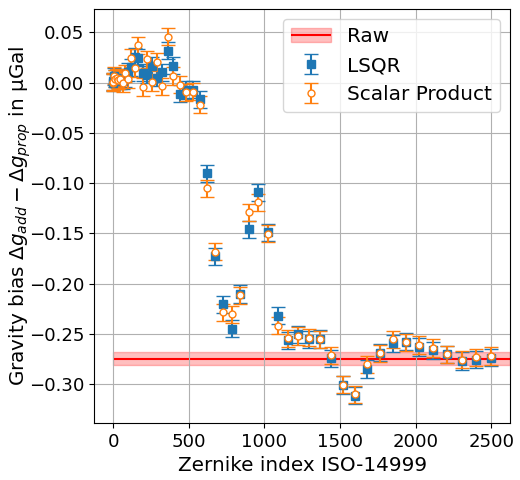}
         \caption{Difference for added and propagated aberrations}
         \label{Fig3b:DecompositionSN1_11mm}
     \end{subfigure}
     \caption{Simulation of the gravimeter with the mirror surface~$1$ shown Figure~\ref{Fig2a:SurfaceMirror1}, for $T=80$~ms, $\sigma_{xy}=300$~$\mu$m and a temperature $\Theta = 2$~$\mu$K. The simulations are done for a decomposition of the mirror surface, either with least mean square \textit{LSQR} or the \textit{scalar product}, on a growing family of Zernike polynomials over a disk of radius $R = 11$~mm, (a) with propagation of the aberrations (errorbars are smaller than the markers). (b) Difference with the case of added aberrations to ideally propagated beam. The analytical points are derived from the 1D integrals \eqref{eq1:WFAddition}-\eqref{eq1:WFPropagation} with the coefficients obtained from the integral decomposition of the mirror surface. The abscissa index corresponds to ISO-14999 numbering \cite{ISO_14999}, which takes into account Zernike polynomials with and without rotation invariance.}
     \label{Fig3:DecompositionSN1_11mm}
\end{figure}

We now examine the impact on the results, of the simulation of the propagation of the aberrations and of their representation on the Zernike polynomial basis, in the case of specific mirrors. For simplicity, the problem is restricted to the case where the mirror, the atomic cloud and the laser beams are coaxially centered. The mirrors surface have been analyzed using a Fizeau interferometer with a $77$~$\mu$m resolution and are shown in Figure~\ref{Fig2:SurfaceMirror}. They are characterized on the central disk of $5$~mm radius by a typical peak-to-valley amplitude of $\approx 9$~nm and an RMS value of $\approx 1$~nm \cite{MirrorSurface_1_2}.
Simulations can be carried out directly using the raw mirror surface data, or the mirror surface can be decomposed on the basis of the Zernike polynomials. The usual method for performing such a decomposition is based on a least mean squares regression, which will be referred as \textit{LSQR} in the following. However, as the Zernike polynomials form an orthogonal family on a reference disk, the decomposition can also be performed by a scalar product consisting of an integration over this disk. Using the mirror surface~$1$ in Figure~\ref{Fig2a:SurfaceMirror1}, simulation with the propagation of the aberrations and the raw data yields a gravity bias $\Delta g^{\text{Raw}}_{\text{prop}} = -8.042(5)$~$\mu$Gal ($1$~$\mu$Gal $=10^{-8}$~m s$^{-2}$), represented by the red line in Figure~\ref{Fig3a:DecompositionSN1_11mm}.
The mirror surface is decomposed over an increasing number of Zernike polynomials, including polynomials without rotational invariance, corresponding to the abscissa of Figure~\ref{Fig3:DecompositionSN1_11mm} \cite{ISO_14999} and gravimeter simulations are performed with these decomposed surfaces, both for the least mean squares technique (solid blue squares) and for the scalar product (orange circles) decomposition. In addition, approximate results (green crosses) labeled \textit{1D Integral}, are obtained using the coefficients of the scalar product decomposition together with the integral \eqref{eq1:WFPropagation} and the propagation approximation up to second order outlined in the \textbf{\nameref{Appendix}}. The simulations resulting from the decomposition show several $\mu$Gal of difference with the one based on the raw data at low decomposition order and converge to $\Delta g^{\text{Raw}}_{\text{prop}}$ when considering sufficiently high-order Zernike polynomials, with the exception of the values resulting from the approximations for propagation and 1D integral, which converge to a difference of $-0.794(5)$~$\mu$Gal. The simulations were performed to include all Zernike polynomials with an index less than $2499$, that of $Z_{98}^{0}$, in accordance with ISO-14999 numbering \cite{ISO_14999}. With the mirror surface defined on a disk with radius $R=11$~mm, aberrations up to the characteristic variation length $\frac{R}{99} \approx 110\ \mu$m, of order of the $77\ \mu$m resolution of the raw data, are represented. As shown in Figure~\ref{Fig3a:DecompositionSN1_11mm}, this order is large enough to ensure convergence of the gravity bias. The difference between the results of the two decomposition methods: \textit{LSQR} and the scalar product, is highlighted at the bottom of Figure~\ref{Fig3a:DecompositionSN1_11mm} and reaches the $\mu$Gal level for low-order decomposition. Eventually, the simulations were run a second time with the addition of the aberrations to the ideally propagated beam, with the raw data and both decomposition methods, the difference with the previous results that take propagation into account is shown in Figure~\ref{Fig3b:DecompositionSN1_11mm}. For both methods there is initially no significant difference for decomposition on low-order Zernike polynomials, which corresponds to the case $n \ll \sqrt{\frac{2kR^2}{\Delta z_{1,\text{mir}}}}$ as pointed out in \eqref{eq1:ApproxPropagation}, though the differences converge to that obtained with the use of the raw data: $-0.274(7)$~$\mu$Gal. This implies that, as in the case of the propagation of aberrations, simulations with the addition of the aberrations converge to the value obtained with the raw data when increasing the family of Zernike polynomials used to represent the mirror surface. Simulations with the mirror surface~2 in Figure~\ref{Fig2b:SurfaceMirror2} were carried out and exhibited similar convergence behaviour up to the gravity bias of $-2.836(4)\ \mu$Gal calculated with the raw data.

Thus, to simulate the gravimeter at the $\mu$Gal level, describing optical aberrations with low-order Zernike polynomials is not sufficient, as the result depends on the decomposition order considered and on the decomposition technique used. It is necessary either to consider high-order Zernike decomposition, or to use raw data with sufficiently high resolution, which is determined by the initial size of the cloud and its expansion length through the interferometer, and therefore the propagation of the aberrations during laser beam propagation. Finally, even if it is more convenient to use the raw data from the wavefront sensor to perform a simulation, the Zernike polynomials basis could still be helpful to provide specifications for the optical surfaces used for a particular experiment.

\section{Dependence on the atomic cloud initial position}
\begin{figure}[ht!]
     \centering
     \begin{subfigure}[b]{0.34\textwidth}
         \centering
         \includegraphics[height=5.7cm]{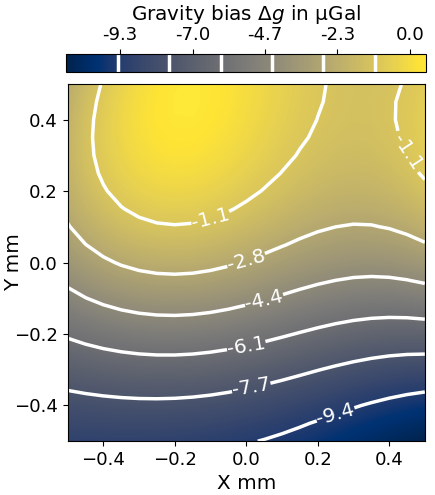}
         \caption{Cloud prepared in a 3D MOT + molasses: $\Theta = 2$~$\mu$K, $\sigma_{xy} = 300$~$\mu$m}
         \label{Fig4a:ChangeInitPos_MOT}
     \end{subfigure}
     \hspace{0.27cm}
     \begin{subfigure}[b]{0.30\textwidth}
         \centering
         \includegraphics[height=5.7cm]{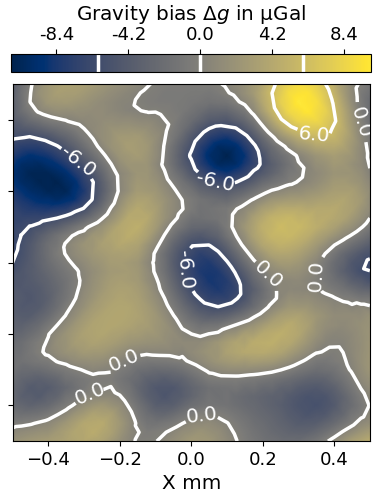}
         \caption{Cloud prepared in a dipolar trap:  $\Theta = 30$~nK, $\sigma_{xy} = 10$~$\mu$m}
         \label{Fig4b:ChangeInitPos_DP}
     \end{subfigure}
     \hspace{0.17cm}
     \begin{subfigure}[b]{0.30\textwidth}
         \centering
         \includegraphics[height=5.7cm]{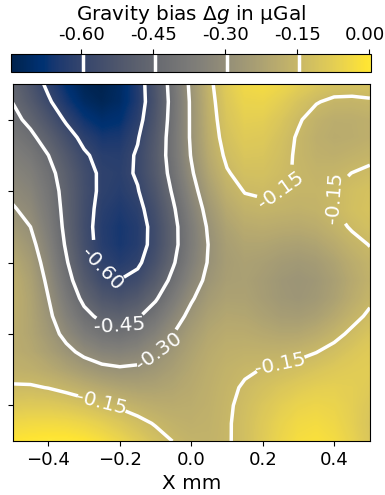}
         \caption{Delta-kicked cloud: $\Theta = 5$~nK, $\sigma_{xy} = 200$~$\mu$m}
         \label{Fig4c:ChangeInitPos_DK}
     \end{subfigure}
     \caption{Simulation of the gravimeter bias in $\mu$Gal with the mirror surface~$2$ represented in Figure~\ref{Fig2b:SurfaceMirror2}, for an interferometer duration $2T = 160$~ms. The initial mean position of the atomic cloud is changed over an area $\left[-500~\mu \text{m}, 500~\mu \text{m} \right] \times \left[-500~\mu \text{m}, 500~\mu \text{m} \right]$ and for three different initial configurations of the atomic cloud.}
     \label{Fig4:ChangeInitPos}
\end{figure}

To compare simulations with experimental results, the initial preparation of the atomic cloud can be modified, for instance a dipole trap can be used after the magneto-optical trap (MOT) and molasses sequence to lower the initial cloud temperature \cite{karcherImprovingAccuracyAtom2018}, and delta-kick collimation technique could be used to obtain a larger cloud with a lower expansion temperature \cite{kovachyMatterWaveLensing2015}. However, it is also important to be able to position the atomic cloud relative to the mirror and laser beams, as the spatial dependence of the bias can be significant, as shown below. For sake of simplicity, the mirror and laser beams are considered to be coaxially centered and only the initial mean position of the atomic cloud is modified. In Figure~\ref{Fig4:ChangeInitPos}, simulations are carried out with raw data of mirror surface~2 in Figure~\ref{Fig2b:SurfaceMirror2} for the same interferometer sequence with an interval between Raman pulses of $80$~ms and for different configurations of the initial atomic cloud. In Figure~\ref{Fig4a:ChangeInitPos_MOT}, the cloud is prepared in the usual gravimeter operating scheme (MOT and molasses) with a temperature of $\Theta=2$~$\mu$K and an initial size of $\sigma_{xy} = 300$~$\mu$m. At position $\p{x, y} = \p{0,0}$ the gravity bias is $-2.836(4)\ \mu$Gal, which is lower than the $-8.042(5)\ \mu$Gal obtained with mirror surface~1 Figure~\ref{Fig2a:SurfaceMirror1}. In addition, while the gravity bias gradient is of the order of $10$~$\mu$Gal~mm$^{-1}$ at the center of the zone of interest, the area for $-500$~$\mu$m~$\leq x \leq$ ~$500$~$\mu$m and $200$~$\mu$m~$\leq y \leq$ ~$500$~$\mu$m exhibits a typical gradient of the order of $1$~$\mu$Gal~mm$^{-1}$, which would enhance the stability of the apparatus with respect to intensity fluctuations of the MOT laser beams that cause fluctuations in initial position \cite{louchet-chauvetInfluenceTransverseMotion2011}. In comparison, the same simulation performed with the mirror surface~1 in Figure~\ref{Fig2a:SurfaceMirror1} shows a gravity bias gradient of the order of $10\ \mu$Gal~mm$^{-1}$ over most of the area under consideration. Consequently, the study of surface~2 Figure~\ref{Fig2b:SurfaceMirror2} with different initial configurations of the atomic cloud has been privileged. In Figure~\ref{Fig4b:ChangeInitPos_DP}, the cloud is prepared in a dipole trap with a temperature of $\Theta=30$~nK and an initial size of $\sigma_{xy} = 10$~$\mu$m, and Figure~\ref{Fig4c:ChangeInitPos_DK} shows simulation for a delta-kicked cloud with a temperature of $\Theta=5$~nK and an initial size of $\sigma_{xy} = 200$~$\mu$m.

For the cloud in the MOT or delta-kick configuration in Figures~\ref{Fig4a:ChangeInitPos_MOT} and \ref{Fig4c:ChangeInitPos_DK}, the typical variation lengths of the gravity bias correspond to the initial cloud size, which is larger than the $77$~$\mu$m mirror surface resolution. Though, in the dipole trap configuration in Figure~\ref{Fig4b:ChangeInitPos_DP}, where the cloud size at the first Raman pulse is roughly $ 30$~$\mu$m, the typical variation length is close to the mirror resolution. In the latter case, simulation results are limited by the initial resolution of the mirror surface data. This is consistent with the simulations of the cloud prepared in a dipole trap at a temperature of $\Theta = 30$~nK shown in Figure~\ref{Fig1:DipolarSimulation}, as the surface defined by Zernike polynomial $Z_{148}^0$ which has a typical variation length of $\frac{R}{n+1} \approx 90\ \mu$m still exhibits large contribution to the phase bias. 

In the delta-kicked cloud configuration of Figure~\ref{Fig4c:ChangeInitPos_DK}, the gradient is at maximum of the order of $1$~$\mu$Gal~mm$^{-1}$ on the overall represented area, and the gravity bias is closer to the ideal case of the Gaussian beam reflected on a perfectly flat mirror, which yields a bias of $-0.21(1)$~$\mu$Gal. This result is consistent with equation \eqref{eq1:ApproxPropagation}, as the dependence of the interferometric phase shift over the initial size is a Gaussian function, increasing the size of the source should result in bias closer to the ideal case.

\section{Conclusion}
This paper demonstrates that accurately simulating wavefront aberrations in a cold-atom gravimeter at the mrad level requires more than just low-order Zernike polynomials to represent the retro-reflecting optics, as this will lead to errors of this order depending on the decomposition order and the decomposition method chosen. In addition, for the distance considered between the retro-reflecting optics and the positions of the Raman pulses, and the spatial frequencies of the aberrations considered, it is necessary to take into account the propagation of the aberrations and not simply add them to the beam propagated in the ideal case. The study of the dependence of the bias on the initial position of the atomic cloud shows that it would be necessary in a MOT configuration to be able to position the cloud relative to the mirror and laser beams with an accuracy of the order of $100$~$\mu$m to reach the mrad level. Eventually, the use of a delta-kicked atomic source could reduce both the interferometer phase shift caused by wavefront aberrations and the sensitivity associated with the initial position of the atomic cloud, as the resulting bias is closer to the ideal case of a beam reflected on a flawless mirror.

This simulation is not exhaustive and focuses on the effect of wavefront aberrations, it could be developed in order to take into account additional features, such as the Coriolis effect and the two-photon light shift which are the next largest contributors to the inaccuracy budget \cite{louchet-chauvetInfluenceTransverseMotion2011, karcherImprovingAccuracyAtom2018}, as well as the residual differential light shift caused by the intensity fluctuations resulting from the propagation of the aberrations and the extra-photon recoil. Moreover, the characterization of the beam at the collimator output and of the different optics on the beam path, together with the use of an atomic distribution and detection response closer to experimental conditions \cite{farahEffectiveVelocityDsitribution2014}, could improve the accuracy of this simulation. In addition, the simulations are based on data acquired with a Fizeau interferometer having a spatial resolution of $77$~$\mu$m, enabling simulations of the device to be carried out for atomic clouds whose typical transverse size is greater than this resolution. For reasons of optical engineering and speed, these devices generally only allow the measurement to be projected onto the first Zernike polynomials. Consequently, this artificially reduces the resolution of the resulting surface and smooths it, introducing errors into the simulation of the interferometer phase compared to using raw data or an extensive decomposition up to high order Zernike polynomials with typical variation size that matches the resolution, as demonstrated in this paper. Furthermore, it is possible, by combining measurements from different surface characterization tools, such as a Fizeau interferometer and a coherence scanning interferometer allowing $\mu$m resolution~\cite{HeOpticalSurfaceCharac2013}, to achieve a resolution lower than $10\ \mu$m that would enable accurate simulation with an atomic cloud having a typical transverse size of a few tens of $\mu$m, for instance after a cooling step in a dipolar trap. However, accounting for these higher spatial frequency components would require a more numerical intensive simulation of the laser field propagation \cite{kozackiNumericalErrorsDiffraction2008}, as well as to consider the longitudinal distribution of the cloud and to calculate the field along this longitudinal direction instead of simply using the mean position of the classical path, given that the characteristic variation length due to propagation will be of the order of the separation of the wavepackets.

Thanks to these simulations and their future comparisons with the direct measurement with the cold-atom gravimeter, we hope to be able to improve the characterization of our device to determine the targeted correction at the $\mu$Gal level. More generally, the evaluation of wavefront aberrations will also be necessary to achieve better accuracy and complete the error budget of other atom interferometry inertial sensors, such as accelerometers embarked in space missions, which target beyond state-of-the-art performances~\cite{levequeCarioqaDefinitionQuantumPathfinder2022}, or differential sensors, such as gradiometers, for which common-mode aberration rejection is limited by the propagation effect considered here. The methodology presented in this paper will be useful for carrying out such a task, adapting the parameters to the experiment under consideration.

\section*{Appendix: Propagation of aberrated beam}\label{Appendix}
The radial part of Zernike polynomials are specific cases of Jacobi polynomials \cite{niuZernikePolynomialsApplications2022}
\begin{equation}
    R_n^m\p{ x} = \p{-1}^{\frac{n-m}{2}} x^m P_{\frac{n-m}{2}}^{\p{m,0}} \p{1-2x^2} \text{ with } |x| \leq 1.
\end{equation}
Using the Jacobi polynomials convergence property \cite{baratellaJacobiUniform1988} and for $|x| \ll 1$ the radial Zernike polynomials can be approximated by a Bessel function
\begin{equation}\label{eqA:BesselApprox}
    R_n^m\p{ x} \approx  \p{-1}^{\frac{n-m}{2}} \frac{\p{\frac{n+m}{2}}!}{\p{\frac{n-m}{2}}! \p{\frac{n+1}{2}}^m } J_m\p{ \p{n+1} x}.
\end{equation}
Considering a beam with Gaussian amplitude and a wavefront described by a Zernike polynomial with its radial part replaced by the Bessel approximation \eqref{eqA:BesselApprox}, would lead to first order in amplitude of wavefront aberration to 
a generalized Bessel Gauss expression \cite{baginiGeneralizedBesselGauss1996}. To have an analytical expression up to second order, a beam with a flat intensity and a wavefront described by a Zernike polynomial with rotational invariance is considered
\begin{equation}\label{eqA:FieldApprox}
    \begin{split}
        U\p{\rho, 0} & = A e^{i\alpha Z_n^0\p{\frac{\rho}{R}}} \approx A e^{i\alpha_n J_0\p{\frac{\rho}{\rho_n}}} \\
        & \approx A \left\{ 1 + i\alpha_n J_0\p{\frac{\rho}{\rho_n}} - \frac{\alpha_n^2}{2} J_0^2\p{\frac{\rho}{\rho_n}}\right\}
    \end{split}
\end{equation}
With $\rho_n = \frac{R}{n+1}$. The first two terms can be propagated by solving the paraxial equation using the closure relation $\int_0^\infty u J_0\p{au} J_0\p{bu} du = \frac{\delta\p{a=b}}{a}$ and the Hankel transform.
Eventually, using a continuity argument at $z=0$, the Bessel function is replaced by the initial Zernike polynomial
\begin{equation}\label{eqA:FirstOrder}
    U^{\p{1}}\p{\rho, z} = A e^{-i kz}  \left\{ 1 + i\alpha e^{i\frac{z}{2k\rho_n^2}} Z_n^0\p{\frac{\rho}{R}}\right\}.
\end{equation}
This first order calculation can be generalized to Zernike polynomials without rotational invariance, by replacing $Z_n^0$ by $Z_n^m$ in equation \eqref{eqA:FirstOrder}. To treat the second order term of expression \eqref{eqA:FieldApprox}, the integral of the triple product of $0^{\text{th}}$ order Bessel function \cite{gervoisTripleBesselIntegral1984} is used
\begin{equation}
    \begin{split}
        &\int_0^\infty \rho J_0\p{a\rho} J_0\p{b\rho} J_0\p{c\rho} d\rho = \frac{1}{2\pi \Delta} \delta_{\Delta > 0} \\
        &\Delta^2 = s \p{s-a} \p{s-b} \p{s-c} \ \text{with} \ s = \frac{1}{2}\p{a+b+c}.
    \end{split}
\end{equation}
Then performing a change of variable $u = \arcsin{\p{\frac{K}{2\rho_n}}}$, with $K$ the coordinate in reciprocal space, the propagated expression of the second order term is derived using the integral $\int_0^{\frac{\pi}{2}} J_0\p{2\gamma \sin{ \theta}} \cos{\p{2\mu \theta}} = \frac{\pi}{2} J_\mu^2\p{\gamma}$ \cite{abramowitzHandbookMathematicalFunctions1972}, 
\begin{equation}
    \begin{split}\label{eqA:SecondCorrection}
        U^{\p{2}}\p{\rho, z} =& - \frac{\alpha^2}{2} A e^{-i kz} e^{i\frac{z}{k \rho_n^2}} \times \left\{J_0\p{\frac{ z}{k\rho_n^2}} J_0^2\p{\frac{\rho}{\rho_n}} \right.\\
        &\ \ \ + 2 \sum_{p=1}^\infty \p{-1}^p J_{2p} \p{\frac{ z}{k\rho_n^2}} J_{2p}^2\p{\frac{\rho}{\rho_n}} \\ 
        &\ \ \ \left. - 2 i \sum_{p=0}^\infty \p{-1}^p J_{2p+1} \p{\frac{ z}{k\rho_n^2}} J_{2p+1}^2\p{\frac{\rho}{\rho_n} } \right\}.
    \end{split}
\end{equation}
Again, by continuity at $z=0$, the Bessel function $J_0\p{\frac{\rho}{\rho_n}}$ can be replaced by the initial Zernike polynomial $Z_n^0\p{\frac{\rho}{R}}$.

\begin{backmatter}
\bmsection{Funding} The authors acknowledge the support from a government grant managed by the Agence Nationale de la Recherche under the Plan France 2030 with the reference “ANR-22-PETQ-0005”, and the Agence Nationale de la Recherche for its financial support of the TONICS project "ANR-21-CE47-0017". This work has been supported by R\' egion Ile-de-France in the framework of DIM SIRTEQ. 

\bmsection{Acknowledgment}
The authors would like to thank Robin Corgier for fruitful discussions and Yann Balland for his digital assistance.

\bmsection{Disclosures}
The authors declare no conflicts of interest.










\bmsection{Data availability} Data underlying the results presented in this paper are not publicly available at this time but may be obtained from the authors upon reasonable request.







\end{backmatter}



\bibliography{sample}

\begin{thebibliography}{10}
\newcommand{\enquote}[1]{``#1''}

\bibitem{riehleOpticalRamseySpectroscopy1991}
F.~Riehle, T.~Kisters, A.~Witte, \emph{et~al.}, \enquote{Optical ramsey spectroscopy in a rotating frame: Sagnac effect in a matter-wave interferometer,} {\protect\JournalTitle{Phys. Rev. Lett.}} \textbf{67}, 177--180 (1991).

\bibitem{kasevichAtomicInterferometry1991}
M.~Kasevich and S.~Chu, \enquote{Atomic interferometry using stimulated raman transitions,} {\protect\JournalTitle{Phys. Rev. Lett.}} \textbf{67}, 181--184 (1991).

\bibitem{bongsTakingAIQuantumSensors2019}
K.~Bongs, M.~Holynski, J.~Vovrosh, \emph{et~al.}, \enquote{Taking atom interferometric quantum sensors from the laboratory to real-world applications,} {\protect\JournalTitle{Nature Reviews Physics}} \textbf{1}, 731--739 (2019).

\bibitem{geigerHighAccuracyInertialMeasurements2020}
R.~Geiger, A.~Landragin, S.~Merlet, and F.~Pereira Dos~Santos, \enquote{High-accuracy inertial measurements with cold-atom sensors,} {\protect\JournalTitle{AVS Quantum Science}} \textbf{2} (2020).

\bibitem{louchet-chauvetInfluenceTransverseMotion2011}
A.~{Louchet-Chauvet}, T.~Farah, Q.~Bodart, \emph{et~al.}, \enquote{The influence of transverse motion within an atomic gravimeter,} {\protect\JournalTitle{New Journal of Physics}} \textbf{13}, 065025 (2011).

\bibitem{GautierAccurateMeasurementSagnac2022}
R.~Gautier, M.~Guessoum, L.~A. Sidorenkov, \emph{et~al.}, \enquote{Accurate measurement of the sagnac effect for matter waves,} {\protect\JournalTitle{Science Advances}} \textbf{8}, eabn8009 (2022).

\bibitem{janvierCompactDifferentialGravimeter2022}
C.~Janvier, V.~M\'enoret, B.~Desruelle, \emph{et~al.}, \enquote{Compact differential gravimeter at the quantum projection-noise limit,} {\protect\JournalTitle{Phys. Rev. A}} \textbf{105}, 022801 (2022).

\bibitem{geigerAirborneMatterWave2011}
R.~Geiger, V.~M{\'e}noret, G.~Stern, \emph{et~al.}, \enquote{Detecting inertial effects with airborne matter-wave interferometry,} {\protect\JournalTitle{Nature Communications}} \textbf{2}, 474 (2011).

\bibitem{bidelMarineGravimetry2018}
Y.~Bidel, N.~Zahzam, C.~Blanchard, \emph{et~al.}, \enquote{Absolute marine gravimetry with matter-wave interferometry,} {\protect\JournalTitle{Nature Communications}} \textbf{9}, 627 (2018).

\bibitem{antoni-micollierDetectingVolcano2022}
L.~Antoni-Micollier, D.~Carbone, V.~Ménoret, \emph{et~al.}, \enquote{Detecting volcano-related underground mass changes with a quantum gravimeter,} {\protect\JournalTitle{Geophysical Research Letters}} \textbf{49}, e2022GL097814 (2022).

\bibitem{gunterMobileFieldMeasurements2024}
A.~Güntner, M.~Reich, J.~Glässel, \emph{et~al.}, \enquote{Mobile field measurements with a quantum gravimeter: Technical setup and performance,} {\protect\JournalTitle{IEEE Instrumentation \& Measurement Magazine}} \textbf{27}, 53--59 (2024).

\bibitem{bouchendiraNewDeterminationFineConstant2011}
R.~Bouchendira, P.~Clad\'e, S.~Guellati-Kh\'elifa, \emph{et~al.}, \enquote{New determination of the fine structure constant and test of the quantum electrodynamics,} {\protect\JournalTitle{Phys. Rev. Lett.}} \textbf{106}, 080801 (2011).

\bibitem{ballandQuectonewtonLocalForceSensor2023}
Y.~Balland, L.~Absil, and F.~Pereira Dos~Santos, \enquote{Quectonewton local force sensor,} {\protect\JournalTitle{Phys. Rev. Lett.}} \textbf{133}, 113403 (2024).

\bibitem{BeckerSpaceBornBEC2018}
D.~Becker, M.~D. Lachmann, S.~T. Seidel, \emph{et~al.}, \enquote{Space-borne bose--einstein condensation for precision interferometry,} {\protect\JournalTitle{Nature}} \textbf{562}, 391--395 (2018).

\bibitem{AvelineObservingBECOrbit2020}
D.~C. Aveline, J.~R. Williams, E.~R. Elliott, \emph{et~al.}, \enquote{Observation of bose--einstein condensates in an earth-orbiting research lab,} {\protect\JournalTitle{Nature}} \textbf{582}, 193--197 (2020).

\bibitem{liRealizationCAGyro2024}
J.~Li, X.~Chen, D.~Zhang, \emph{et~al.}, \enquote{Realization of a cold atom gyroscope in space,}  (2024). ArXiv 2405.20659.

\bibitem{levequeCarioqaDefinitionQuantumPathfinder2022}
T.~Lévèque, C.~Fallet, J.~Lefebve, \emph{et~al.}, \enquote{Carioqa: Definition of a quantum pathfinder mission,}  (2022). ArXiv 2211.01215.

\bibitem{abendRecentDevelopmentsQT2023}
S.~Abend, B.~Allard, A.~S. Arnold, \emph{et~al.}, \enquote{{Technology roadmap for cold-atoms based quantum inertial sensor in space},} {\protect\JournalTitle{AVS Quantum Science}} \textbf{5}, 019201 (2023).

\bibitem{struckmannPlatformEnvironmentRequirements2024}
C.~Struckmann, R.~Corgier, S.~Loriani, \emph{et~al.}, \enquote{Platform and environment requirements of a satellite quantum test of the weak equivalence principle at the ${10}^{\ensuremath{-}17}$ level,} {\protect\JournalTitle{Phys. Rev. D}} \textbf{109}, 064010 (2024).

\bibitem{badeObservationExtraPhotonRecoil2018}
S.~Bade, L.~Djadaojee, M.~Andia, \emph{et~al.}, \enquote{Observation of extra photon recoil in a distorted optical field,} {\protect\JournalTitle{Phys. Rev. Lett.}} \textbf{121}, 073603 (2018).

\bibitem{cervantesEffectAperture2024}
J.~M. Cervantes and E.~Gomez, \enquote{Effect of an aperture in atomic gravimetry,} {\protect\JournalTitle{J. Opt. Soc. Am. A}} \textbf{41}, 881--891 (2024).

\bibitem{schkolnikEffectWavefrontAberrations2015}
V.~Schkolnik, B.~Leykauf, M.~Hauth, \emph{et~al.}, \enquote{The effect of wavefront aberrations in atom interferometry,} {\protect\JournalTitle{Applied Physics B}} \textbf{120}, 311--316 (2015).

\bibitem{zhouObservingEffectWavefrontAberrations2016}
M.-K. Zhou, Q.~Luo, L.-l. Chen, \emph{et~al.}, \enquote{Observing the effect of wave-front aberrations in an atom interferometer by modulating the diameter of raman beams,} {\protect\JournalTitle{Phys. Rev. A}} \textbf{93}, 043610 (2016).

\bibitem{karcherImprovingAccuracyAtom2018}
R.~Karcher, A.~Imanaliev, S.~Merlet, and F.~P.~D. Santos, \enquote{Improving the accuracy of atom interferometers with ultracold sources,} {\protect\JournalTitle{New Journal of Physics}} \textbf{20}, 113041 (2018).

\bibitem{trimecheActiveControlLaser2017}
A.~Trimeche, M.~Langlois, S.~Merlet, and F.~Pereira Dos~Santos, \enquote{Active {{Control}} of {{Laser Wavefronts}} in {{Atom Interferometers}},} {\protect\JournalTitle{Physical Review Applied}} \textbf{7}, 034016 (2017).

\bibitem{Zernike_low_order}
As the typical variation length of the Zernike polynomial $Z^0_n$ is $\rho_n = \frac{R}{n+1}$, with $R$ the radius over which the Zernike polynomial is defined, see \textbf{\nameref{Appendix}}, we refer to Zernike polynomials as being low order, when their typical variation length is larger than the typical transverse size of the atomic cloud during the experiment.

\bibitem{Lan2012Influence}
S.-Y. Lan, P.-C. Kuan, B.~Estey, \emph{et~al.}, \enquote{Influence of the coriolis force in atom interferometry,} {\protect\JournalTitle{Phys. Rev. Lett.}} \textbf{108}, 090402 (2012).

\bibitem{kozackiNumericalErrorsDiffraction2008}
T.~Kozacki, \enquote{Numerical errors of diffraction computing using plane wave spectrum decomposition,} {\protect\JournalTitle{Optics Communications}} \textbf{281}, 4219--4223 (2008).

\bibitem{bornwolfPrinciplesOptics2019}
M.~Born and E.~Wolf, \emph{The diffraction theory of aberrations} (Cambridge University Press, 2019), pp. 517–--553.

\bibitem{niuZernikePolynomialsApplications2022}
K.~Niu and C.~Tian, \enquote{Zernike polynomials and their applications,} {\protect\JournalTitle{Journal of Optics}} \textbf{24}, 123001 (2022).

\bibitem{kovachyMatterWaveLensing2015}
T.~Kovachy, J.~M. Hogan, A.~Sugarbaker, \emph{et~al.}, \enquote{Matter wave lensing to picokelvin temperatures,} {\protect\JournalTitle{Phys. Rev. Lett.}} \textbf{114}, 143004 (2015).

\bibitem{GeneralizationPhaseBias}
The integral of the Zernike polynomial $Z_n^m\left(\frac{\rho}{R}, \theta\right)$ with a gaussian distribution centered on $\vec{r}_0 = \rho_0 \left(\cos{\left( \theta_0 \right)}, \sin{\left( \theta_0 \right)}\right)$ and of standard deviation $\sigma_\rho$ results in $ \exp{\left(-\frac{\sigma_\rho^2}{2 \rho_n^2}\right)} Z_n^m\left( \frac{\rho_0}{R}, \theta_0 \right)$, provided that the Zernike polynomial can be replaced by its Bessel function approximation, and with $\rho_n = \frac{\rho}{n+1}$. Thus, the equations describing the phase shift for the addition or the propagation of aberrations can be generalized by mutilplying them by the value of the Zernike polynomial at the center of the distribution. For the Zernike polynomials $Z_n^0$ this is already done as $\gamma_n = \left(-1\right)^{\frac{n}{2}} \gamma = \gamma Z_n^0\left( 0\right)$.

\bibitem{MirrorSurface_1_2}
While the mirror surface 1 has been analyzed on a disk with radius $R=13.5$~mm, the mirror surface 2 has been analyzed on a disk with radius $R=11$~mm. For simplicity, the mirror surface 1 have been truncated at a radius of $R=11$~mm. The manufacturing specifications for the mirrors surface were the RMS and Peak-to-Valley values on the $5$~mm radius central disk.

\bibitem{ISO_14999}
The ISO-14999 indexing scheme is used to number all Zernike polynomials $Z_n^m$. Starting from $0$, the index $i$ of the Zernike polynomial $Z_n^m$ is given by the relation $i = \p{\frac{n+|m|}{2}+1}^2 - 2|m| - \delta_{m \geq 0}$.

\bibitem{farahEffectiveVelocityDsitribution2014}
T.~Farah, P.~Gillot, B.~Cheng, \emph{et~al.}, \enquote{Effective velocity distribution in an atom gravimeter: Effect of the convolution with the response of the detection,} {\protect\JournalTitle{Phys. Rev. A}} \textbf{90}, 023606 (2014).

\bibitem{HeOpticalSurfaceCharac2013}
L.~He, C.~J. Evans, and A.~Davies, \enquote{Optical surface characterization with the area structure function,} {\protect\JournalTitle{CIRP Annals}} \textbf{62}, 539--542 (2013).

\bibitem{baratellaJacobiUniform1988}
P.~Baratella and L.~Gatteschi, \enquote{The bounds for the error term of an asymptotic approximation of jacobi polynomials,} in \emph{Orthogonal Polynomials and their Applications,}  M.~lfaro, J.~S. Dehesa, F.~J. Marcellan, \emph{et~al.}, eds. (Springer Berlin Heidelberg, Berlin, Heidelberg, 1988), pp. 203--221.

\bibitem{baginiGeneralizedBesselGauss1996}
V.~Bagini, F.~Frezza, M.~Santarsiero, \emph{et~al.}, \enquote{Generalized bessel-gauss beams,} {\protect\JournalTitle{Journal of Modern Optics}} \textbf{43}, 1155--1166 (1996).

\bibitem{gervoisTripleBesselIntegral1984}
A.~Gervois and H.~Navelet, \enquote{{Some integrals involving three Bessel functions when their arguments satisfy the triangle inequalities},} {\protect\JournalTitle{Journal of Mathematical Physics}} \textbf{25}, 3350--3356 (1984).

\bibitem{abramowitzHandbookMathematicalFunctions1972}
M.~Abramowitz and I.~A. Stegun, eds., \emph{Handbook of Mathematical Functions with Formulas, Graphs, and Mathematical Tables} (U.S. Government Printing Office, Washington, DC, USA, 1972), tenth printing ed.

\end{thebibliography}






\end{document}